\documentclass[letter,traditabstract]{aa}
\usepackage{natbib}
\bibpunct{(}{)}{;}{a}{}{,}
\usepackage{graphicx}
\usepackage{revsymb}	
\usepackage{amsmath}	
\usepackage[usenames]{color}	
\usepackage{epstopdf}	
\newcommand{\eq}[1] {Eq.\,(\ref{#1})}

\usepackage{txfonts}

\usepackage[switch]{lineno} 
\nolinenumbers
\modulolinenumbers[5]

\begin{document}
\title{Damping rates of solar-like oscillations across the HR diagram}
\subtitle{Theoretical calculations confronted to CoRoT and {\it Kepler} observations}

\author{K. Belkacem\inst{1} \and  M.A. Dupret\inst{2}  \and F. Baudin\inst{3} \and T. Appourchaux \inst{3} \and J. P. Marques\inst{4,1} \and R. Samadi\inst{1}}

\institute{ 
LESIA, UMR8109, Universit\'e Pierre et Marie Curie, Universit\'e
	Denis Diderot, Obs. de Paris, 92195 Meudon Cedex, France
\and 
Institut dÕAstrophysique et de G\'eophysique, Universit\'e de Li\`ege, All\'ee du 6 Ao\^ut 17-B 4000 Li\`ege, Belgium 
\and
Institut d'Astrophysique Spatiale, CNRS, Universit\'e Paris XI,
   91405 Orsay Cedex, France
   \and
   Georg-August-Universit\"at G\"{o}ttingen, Institut f\"ur Astrophysik, Friedrich-Hund-Platz 1, D-37077 G\"{o}ttingen, Germany
}

   \offprints{K. Belkacem}
   \mail{kevin.belkacem@obspm.fr}
   \date{\today}

  \authorrunning{Belkacem et al.}
  \titlerunning{Damping rates of solar-like oscillations across the HR diagram}

   \abstract{ The space-borne missions CoRoT and {\it Kepler} are providing a rich harvest of high-quality constraints on solar-like pulsators. Among the seismic parameters, mode damping rates remains poorly understood and are thus barely used to infer the physical properties of stars. Nevertheless, thanks to the CoRoT and {\it Kepler} spacecrafts it is now possible to measure damping rates for hundreds of main-sequence and thousands of red-giant stars with unprecedented precision. 
   By using a non-adiabatic pulsation code including a time-dependent convection treatment, we compute damping rates for stellar models that are representative of solar-like pulsators from the main-sequence to the red-giant phase.  This allows us to reproduce the observations of both CoRoT and {\it Kepler}, which validates our modeling of mode damping rates and thus the underlying physical mechanisms included in the modeling.  By considering the perturbations of turbulent pressure and entropy (including the perturbation of the dissipation rate of turbulent energy into heat) by the oscillation in our computation, we succeed in reproducing the observed relation between damping rates and effective temperature. 
   Moreover, we discuss the physical reasons for mode damping rates to scale with effective temperature, as observationally exhibited. Finally, this opens the way for the use of mode damping rates to probe turbulent convection in solar-like stars. }


   \maketitle
   
\section{Introduction}
\label{intro}

The space missions CoRoT and {\it Kepler} are providing accurate observations of solar-like oscillations of  hundreds of main-sequence stars and thousands of red-giant stars. Therefore, we now have access to  seismic properties, such as mode frequencies, amplitudes, and linewidths, for a large and homogeneous sample of stars. Among them, the physical mechanisms governing mode linewidths ($\Gamma$), which is linearly related to mode damping rates ($\eta$) such as $\Gamma=\eta/\pi$, are still poorly understood. Many studies have attempted to model the damping rates of solar modes \citep[e.g.,][]{B92a,Dupret06b,Xiong00}. All those modelings have had difficulties in reproducing the solar observations except when using free parameters, and the underlying physical mechanisms remain unclear \citep[e.g.,][]{Houdek2008}. 

The scaling relation of mode linewidths for other stars thus provides an important additional constraint on the modeling. A usual way to investigate mode parameters adopts scaling relations between  asteroseismic quantities  and stellar  parameters \citep[e.g.][]{Kjeldsen95,Samadi07,Mosser10,Baudin2011a,Baudin2011b,Belkacem2011}. 
For mode linewidths, scaling relations have been investigated only very recently. \cite{Chaplin09} used  ground-based observations and proposed that mode linewidths follow a power-law dependence on effective temperature, \emph{i.e.} $\Gamma \propto T_{\rm eff}^4$ where $\Gamma$ is the mode linewidth and $T_{\rm eff}$ the effective temperature. However, those conclusions were challenged by \cite{Baudin2011a,Baudin2011b}  using a homogeneous sample of stars observed by CoRoT. They found that a unique power-law cannot describe the entire range of effective temperature covered by main-sequence and red-giant stars and  concluded that mode linewidths of main-sequence stars follow a power-law of $T_{\rm eff}^{16 \pm 2}$, while red-giant stars only slightly depend on effective temperature ($T_{\rm eff}^{-0.3 \pm 0.9}$). This result, for main-sequence and sub-giant stars, was later extended by {\it Kepler} observations \citep{Appourchaux2011}. 

From a theoretical point of view, \cite{Chaplin09} predicted a power-law of $\Gamma \propto T_{\rm eff}^4$, which disagrees with CoRoT and {\it Kepler} observations. \cite{Houdek2012} attributed the failure of this theory to a missing physical mechanism and proposed mode scattering as a possible solution. This disagreement emphasizes that our understanding of mode linewidth is still in its infancy and that a first necessary  breakthrough would be to reproduce the  CoRoT and {\it Kepler} observations. Hence, in this letter we adopt the formalism of \cite{MAD05} and \cite{Dupret06a}. This allows us to confront theoretical computations with observations from both the CoRoT and {\it Kepler} spacecrafts, and consequently to understand the strong dependence of mode linewidth on effective temperature and validate the theoretical modeling. 

\section{Computation of theoretical damping rates}
\label{compute_damping}

\subsection{The grid of stellar models}
\label{models}

We used a grid of stellar models for masses between $M=1 \, M_\odot$ and $M=1.4 \,M_\odot$ from the ZAMS to the tip of the red giant branch, which are typical of observed solar-like pulsators. 
The grid was obtained using the  stellar evolution code CESAM2k  \citep{Morel08}. The atmosphere was computed assuming a grey Eddington approximation. 
Convection was included according to the B\"{o}hm-Vitense mixing-length (MLT) formalism, with a mixing-length parameter $\alpha=1.6$.  The initial chemical composition follows \cite{Asplund05}, with an helium mass fraction of $0.2485$. We used the OPAL equation of state
\citep{Rogers96} and opacities \citep{Iglesias96}, 
complemented, at $T < 10^4$ K, by the \cite{Alexander94} opacities. 
We used the NACRE nuclear reaction rates from \cite{Angulo99} except for the 14 N + p reaction, where we used the reaction rates given in \cite{Imbriani04}.

\subsection{The non-adiabatic oscillation code}
\label{non-adiab}

Damping rates were computed using the non-adiabatic 
pulsation code MAD \citep{MAD02}, which includes 
the time-dependent convection (TDC) treatment described in \cite{MAD05}. 
This approach takes into account the role played 
by the variations in the convective flux, the turbulent pressure, and the dissipation rate of 
turbulent kinetic energy. This TDC approach  is a non-local  formulation of  convection based on the \cite{Gabriel96} formalism explained in \cite{Dupret06a,Dupret06b}. In this framework, non-local parameters related to the convective flux and the turbulent pressure are chosen such as in \cite{Dupret06a} so that it fits the solar 3D numerical simulation. 

In addition, it involves a parameter $\beta$, which takes complex values  and enters the closure term of the 
perturbed  energy equation. This parameter was introduced to prevent the occurrence of  non-physical spatial oscillations in the eigenfunctions \cite[see][ for details]{MAD05}. 
To constrain this parameter, we adopt the following strategy: $\beta$ is adjusted so that the frequency of the depression of the damping rates (see Fig.~\ref{fig_theorique}) coincides with $\nu_{\rm max}$, where $\nu_{\rm max}$ is computed by using the linear relation between $\nu_{\rm max}$ and the cut-off frequency  \citep[e.g.,][]{Kjeldsen95}. We note that the bond between $\nu_{\rm max}$ and the frequency of bottom of the second ascending branch of the damping rates is observed for the Sun \citep{Belkacem2011} and more generally for solar-like pulsators \citep{Appourchaux2011}. 

\subsection{Dominant contributions of the damping rates}
\label{contributions}

The integral expression of the damping rates can be written as \citep{MAD05}
\begin{equation}
\label{dampings_radiatif}
\eta = \frac{1}{2 \, \omega I} \int_{0}^{M} \mathcal{I}m \left[ 
\left(\Gamma_3 - 1\right) \, 
\frac{\delta \rho}{\rho_0}^* T_0 \delta S  
+ \frac{\delta \rho}{\rho_0}^* \frac{\delta P_{\rm turb}}{\rho_0} \right] \textrm{d}m \, , 
\end{equation}
where $\omega$ is the mode frequency, $I$ the mode inertia, $(\Gamma_3-1)=(\partial \ln T_0 / \partial \ln \rho_0)_s$, $\delta \rho$ the Lagrangian perturbation of density, $\delta S$ is the Lagrangian perturbations of entropy, $\delta P_{\rm turb}$ the perturbation of turbulent pressure, $T_0$ the mean temperature, $\rho_0$ the mean density, and the star denotes the complex conjugate. 

\begin{figure}
\begin{center}
\includegraphics[height=6cm,width=8cm]{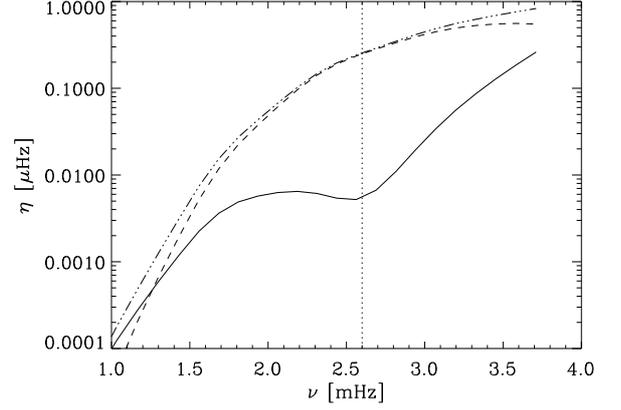}
\caption{Damping rates versus mode frequency, computed for a model of a mass of $M=1.1 \, M_\odot$ and an effective temperature of $T_{\rm eff}=6026 $K on the main sequence. The damping has been computed as described in Sect.~\ref{non-adiab}. The solid line corresponds to the total damping rate, while the dash-dotted line corresponds to the contribution of the perturbation of turbulent pressure and the dashed line to the absolute value of the contribution of the entropy perturbations (see Eq.~\ref{dampings_radiatif}). The vertical dotted line corresponds to $\nu_{\rm max}$.}
\label{fig_theorique}
\end{center}
\end{figure}

The first term of \eq{dampings_radiatif} includes the contributions of the perturbations of the radiative and convective fluxes, as well as the perturbation of the dissipation rate of turbulent kinetic energy into heat. We note that this term corresponds to the non-adiabatic part of the gas pressure perturbation, which is found to be negative (thus a driving contribution). The second term of \eq{dampings_radiatif} represents the perturbation of turbulent pressure. Figure~\ref{fig_theorique} displays the mode damping versus the mode frequency as well as the two contributions expressed in \eq{dampings_radiatif}. It turns out that the contribution of turbulent pressure dominates the damping and is partly compensated by the contribution of entropy. Both contributions have roughly the same order of magnitude, hence the total mode damping is small compared to the absolute values of both the entropy and turbulent pressure contributions. In addition, we note that the depression (or plateau) of the damping rates is the result of the maximum compensation between the two contributions. This is the case for all the models, from the main-sequence to the red-giant phases. 

\section{Comparison with CoRoT and {\it Kepler} observations}
\label{comparaison}

\begin{figure}
\begin{center}
\includegraphics[height=7cm,width=9.3cm]{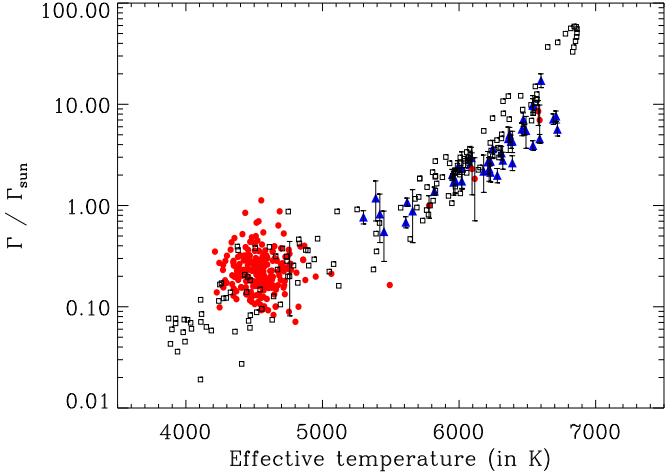}
\caption{Mode linewidths (normalized by the solar value, $\Gamma_{\rm sun} = 0.95 \, \mu$Hz) versus effective temperature. The squared symbols represent theoretical calculations computed as explained in Sects.~\ref{models} and \ref{non-adiab}. The triangles correspond to the observations of main-sequence stars  derived by \cite{Appourchaux2011} from the {\it Kepler} data (with their 3-$\sigma$ error-bars). The dots correspond to the observations of red giants (with $T_{\rm eff} < 5200$ K) and main-sequence  (with $T_{\rm eff} > 5200$ K, with their 3-$\sigma$ error-bars) stars as derived by \cite{Baudin2011a,Baudin2011b} from the CoRoT data. 
}
\label{Kepler_Corot}
\end{center}
\end{figure}

\subsection{Data set}

We considered CoRoT and Kepler stars for which linewidths had been accurately measured. 
We first considered the CoRoT observations described in \cite{Baudin2011a,Baudin2011b}. The measured linewidths for main-sequence stars come from the CoRoT seismological field. More precisely, we used  HD49933 \citep{Benomar09b}, HD181420 \citep{Barban09}, and HD49385 \citep{Deheuvels09}. 
We also used the results for HD50890 \citep{Baudin2011c}, HD181907 \citep{Carrier09}, and the Sun. For CoRoT red-giants,  \cite{Baudin2011a,Baudin2011b}  fitted 235 red giants observed in the exo-field of CoRoT for about 142 days. 

We also considered {\it Kepler} observations of 42 stars acquired over nine months as presented by \cite{Appourchaux2011}. Those stars were analyzed by several groups and mode linewidths at the maximum of height in the power spectrum were derived. We note that \cite{Appourchaux2011} considered only main-sequence stars and sub-giants but not red giants. 

\subsection{Results}

A direct comparison between observations and theoretical computations is provided in Fig.~\ref{Kepler_Corot} for both {\it Kepler} and CoRoT observations. 
For main-sequence stars, there is an overall agreement between the theoretical computations  and both CoRoT and {\it Kepler} observations within the observational error-bars. However, the observed {\it Kepler} linewidths are smaller than predicted by theory and this is particularly the case for stars with high effective temperatures. While this result can be related to modeling deficiencies, there are also several possible observational uncertainties. For instance, the determination of effective temperature is subject to important uncertainties in the estimation of the mode linewidth \citep{Appourchaux2011}.

For red-giant stars, our theoretical computation is in overall agreement with the results derived for CoRoT observations by \cite{Baudin2011a,Baudin2011b}. We note however that possible uncertainties can also come from either the determination of effective temperatures or a bias due to the observation duration and thus the limited frequency resolution in \cite{Baudin2011a,Baudin2011b}. 
An extended investigation of damping rates dedicated to red-giant stars would thus be desirable in a future work, and allow us to draw conclusions about the dependence of effective temperature to damping rates for red-giants. In addition, such a work will benefit from ongoing {\it Kepler} observations that will provide us red giants at low effective temperatures ($T_{\rm eff} <$ 4200\,K). 

This overall agreement with both CoRoT and {\it Kepler} observations demonstrates that the main physical picture is well-reproduced by the modeling. However, an extension of this comparison using sub-giants and high luminosity red-giant stars observed by {\it Kepler} is desirable in the future to ensure that we have a homogeneous sample of stars across a large range of effective temperatures. 

\section{Discussions}
\label{theory}

We have discussed the strong dependence of mode damping rates on effective temperature and shown that it can be understood by simple arguments. Secondly, we have investigated the possible origin of a different scaling for the damping rates of main-sequence and red-giant stars. 

\subsection{Mode linewidth versus effective temperature}
\label{explanation}

\begin{figure}
\begin{center}
\includegraphics[height=6cm,width=8cm]{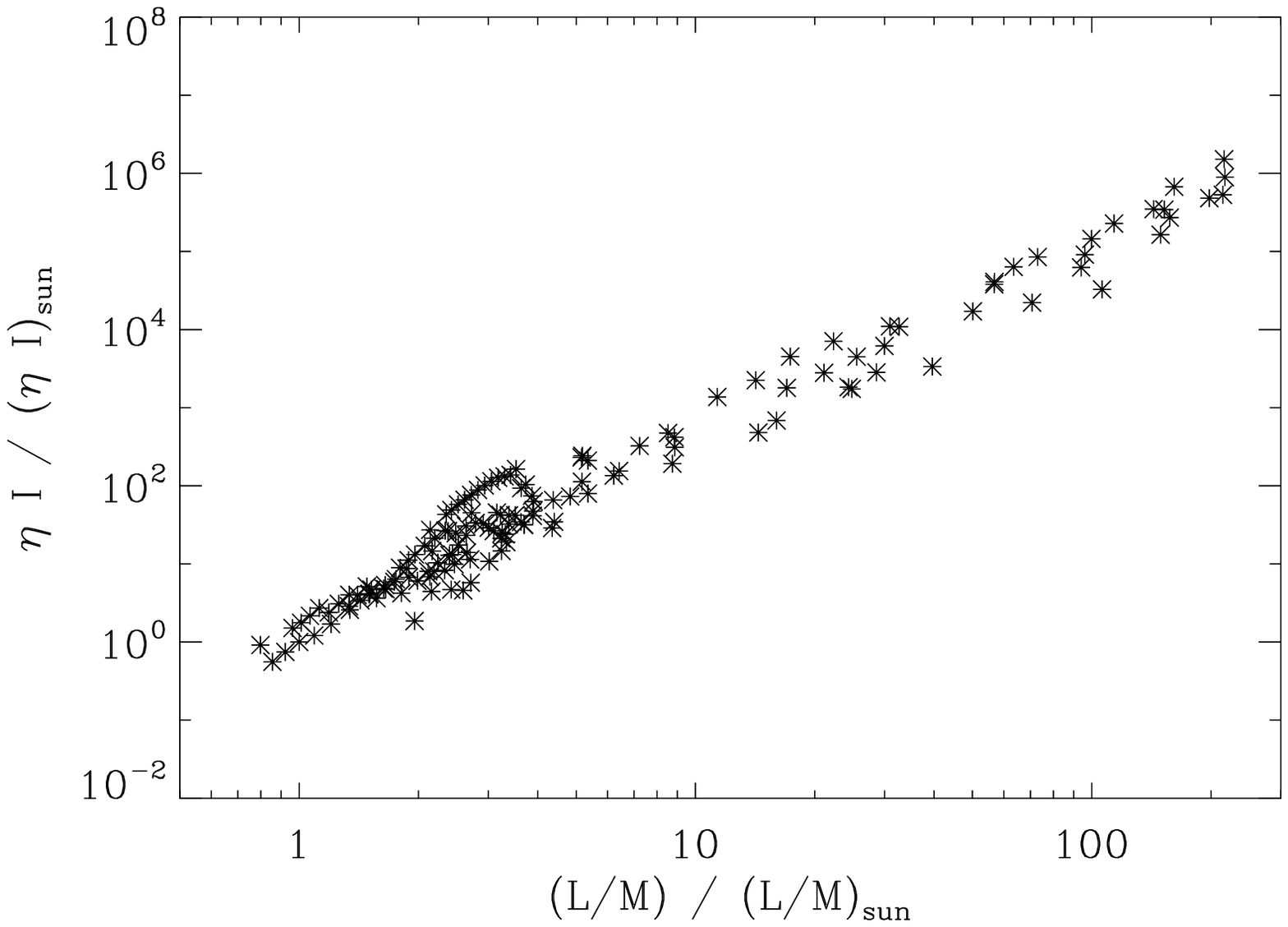}
\includegraphics[height=6cm,width=8cm]{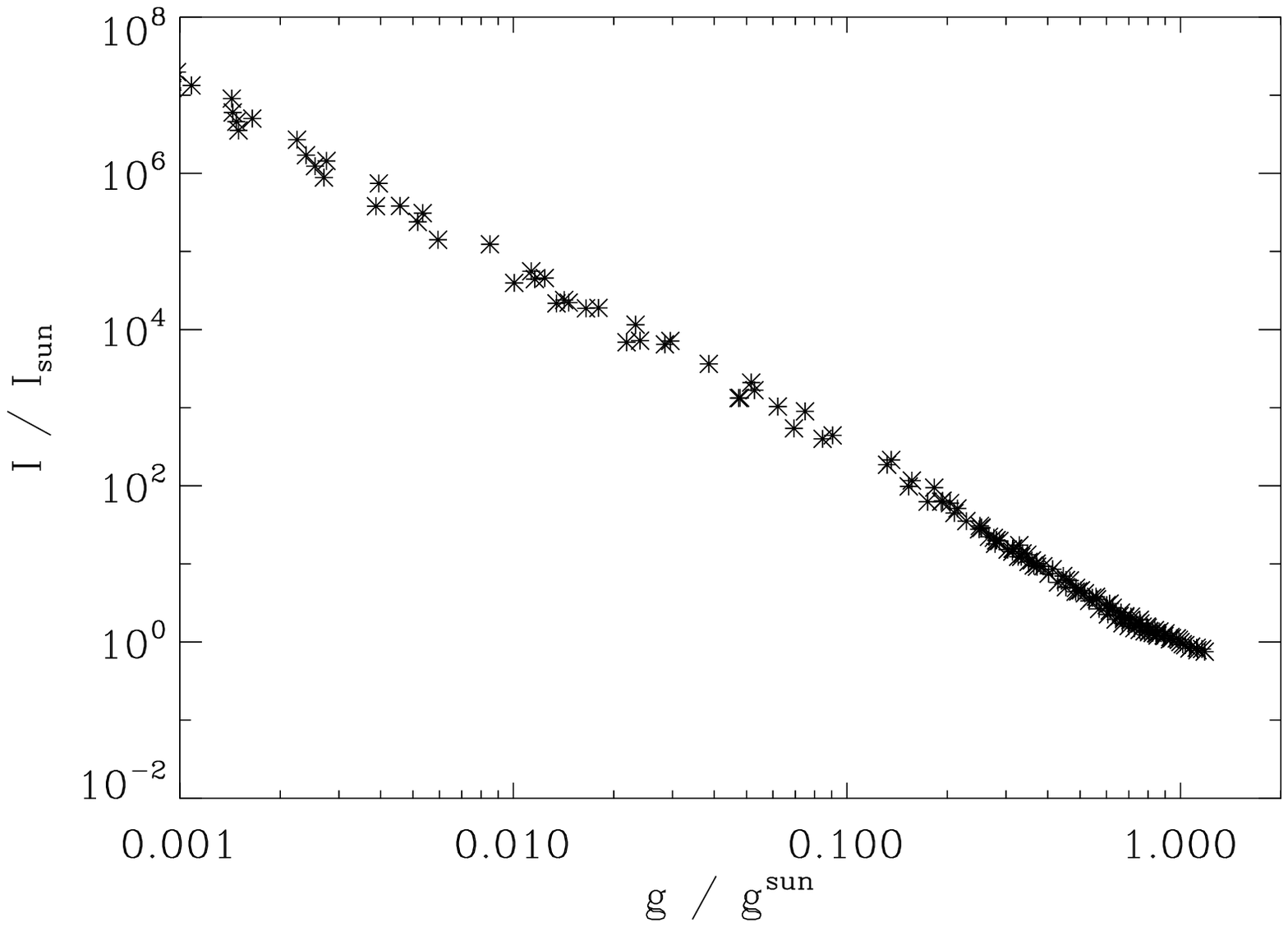}
\caption{{\it Top:} Product $\eta \times I$ (see \eq{dampings_radiatif} for details) as a function of the ratio $L/M$ (where $L$ and $M$ are the luminosity and mass, respectively). {\it Bottom:} Mode inertia versus surface gravity $g$.}
\label{etaI_inertie}
\end{center}
\end{figure}

We start from \eq{dampings_radiatif} and note that there are obviously two important quantities, the work integral and the mode inertia. The first depends on the phase difference between the mode compression (perturbation of density) and  perturbation of pressure (gas and turbulent pressure). Hence, it is the non-adiabatic part of pressure fluctuations ($\delta S / c_v$) that mainly determines the mode damping.   According to the mode energy equation \cite[see, for instance, Eq. (A2) of ][]{Belkacem2011}, one can expect that the work integral dimensionally scales as the ratio $L/M$. 
This is confirmed by Fig.~\ref{etaI_inertie} (top), which shows that the work integral follows
\begin{equation}
\label{etaI}
\eta \, I \propto \left( \frac{L}{M} \right)^{2.7} \, .
\end{equation}
In contrast, mode inertia ($I$) does not depend on mode energy leakage but on the star's static structure and more precisely on the properties of the uppermost layers. Hence, one can expect mode inertia to scale with the surface gravity\footnote{Note that mode inertia also scales with the dynamical timescale $\sqrt(GM/R^3)$ with almost the same dispersion as for the surface gravity.} as in the case shown by Fig.~\ref{etaI_inertie} (bottom). More precisely, 
\begin{equation}
\label{Inertie}
I \propto g^{-2.4} \, .
\end{equation}
Now using \eq{etaI} and \eq{Inertie} and further noting that $L/M \propto T_{\rm eff}^4/g$, one easily finds that mode damping depends mainly on effective temperature, such that 
\begin{equation}
\eta \propto T_{\rm eff}^{10.8} \; g^{-0.3} \, .
\end{equation}
Such a crude analysis is unable to reproduce the precise shape of the mode line-width with effective temperature. However, it allows us to explains qualitatively the strong dependence of mode damping rates on effective temperature. It turns out that the dependence on the effective temperature is a result of a compensation between the work integral and mode inertia. 

\subsection{Main-sequence and red-giants stars: two physical regimes?}
\label{regimes}

Several authors \citep{Ando2010,Baudin2011a,Baudin2011b} have shown by using both ground-based observations and photometric observations from space that there is an apparent change in the behavior of the  mode damping rates between main-sequence and red-giant stars. They therefore suggested that there is a switch between two different physical regimes. However, our theoretical results suggest that the physical mechanisms for mode damping remain the same from main-sequence to red-giant stars. 

\begin{figure}
\begin{center}
\includegraphics[height=6cm,width=8cm]{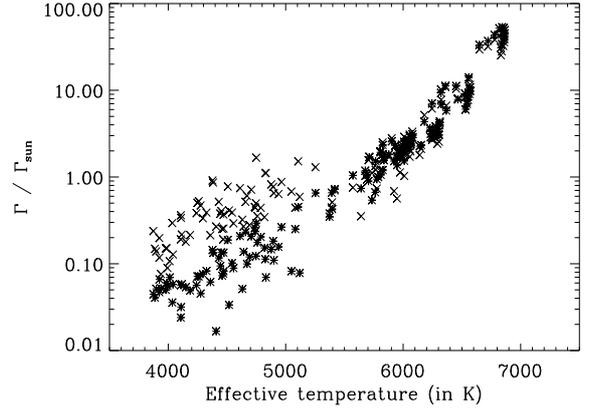}
\caption{As for Fig.~\ref{Kepler_Corot}. Double cross symbols correspond, as described in Sect.~\ref{comparaison}, to damping rates computed for the closest mode in frequency to the frequency of the maximum  $\nu_{\rm max}$, minus one radial order. The crosses correspond to the same calculation but adding one radial order for the selection of the frequency of the maximum. This procedure is intended to mimic the effect of  possible biases in the observational determination of $\nu_{\rm max}$. }
\label{eta_var_max}
\end{center}
\end{figure}

However, the power-law of mode damping with effective temperature is very sensitive to the way $\nu_{\rm max}$ is selected. This is illustrated by Fig.~\ref{eta_var_max}. For instance, a bias in selecting $\nu_{\rm max}$ can arise because in red giants some non-radial modes have almost the same widths as radial ones \citep{Dupret09}. For main-sequence stars, the mode density is important so that an error of, say,  one radial order in the selection of $\nu_{\rm max}$ has negligible effects. In contrast, for red giants, this shift can lead to the selection of a mode outside the depression of the damping rates. Hence, we conclude that the slope of mode linewidth, for red-giants, with effective temperature is sensitive to the way in which the frequency of the maximum height is derived in each power spectrum. In turn, it could also explain why \cite{Appourchaux2011} found different results from their fitting when the frequency of the maximum height or amplitude was selected. 

\section{Conclusions}
\label{conclu}

In contrast to previous work, we have shown that the theoretical computation of mode damping rates agree  with both {\it Kepler} and CoRoT observations from main-sequence to red-giant stars. This result suggests that the main physical picture used to model mode damping successfully reproduces the observations. Moreover, we have been able to understand the strong dependence of mode damping rates on effective temperature by pointing out that it is the result of a compensation between the work integral and mode inertia. 

In contradiction again to previous results our theoretical computation of mode damping rates with the same physics description are in agreement with both Kepler and CoRoT observations from main-sequence (strong relation with $T_{\rm eff}$) to red-giant stars, there being a large dispersion in the temperature range 4500$\pm$250\,K. Our results at lower temperature ($T_{\rm eff} <$ 4200\,K) again show that there is a strong dependence of the damping on $T_{\rm eff}$. Damping measurements for stars in this range using for example the ongoing {\it Kepler} mission will demonstrate whether  our description is valid for this range and put tighter constraints on the relation for red giants. 


\end{document}